\documentstyle[epsfig]{aipproc}
\begin{document}

\title{Parity Violating Measurements of Neutron Densities and Nuclear Structure} 

\author{C. J. Horowitz}

\address{Dept. of Physics and Nuclear Theory Center,
Indiana University, Bloomington, IN 47405 USA \\ E-mail: charlie@iucf.indiana.edu}

\maketitle

\begin{abstract}
Parity violating electron nucleus scattering is a clean
and powerful tool for measuring the spatial distributions of 
neutrons in nuclei with unprecedented accuracy.
Parity violation arises from the interference of electromagnetic
and weak neutral amplitudes, and the $Z^0$ of the 
Standard Model couples primarily to neutrons at low $Q^2$.
Experiments are now feasible at existing facilities.  
We show that theoretical corrections are either small or
well understood, which makes the interpretation clean. A neutron density measurement may have many implications for nuclear structure, atomic parity nonconservation experiments, and the structure of neutron stars.
\end{abstract}

\section*{Introduction}

The size of a heavy nucleus is one of its most basic properties.  However,
because of a neutron skin of uncertain thickness, the size does not follow from
measured charge radii and is relatively poorly known.  For example, the root
mean square neutron radius in $^{208}$Pb, $R_n$ is thought to be about 0.25 fm
larger then the proton radius $R_p\approx 5.45$ fm.  An accurate measurement of
$R_n$ would provide the first clean observation of the neutron skin in a stable
heavy nucleus.  This is thought to be an important feature of all heavy nuclei.

Ground state charge densities have been determined from elastic electron
scattering, see for example ref.\cite{Sick}.  Because the densities are both
accurate and model independent they have had a great and lasting impact on
nuclear physics.  They are, quite literally, our modern picture of the nucleus.

In this paper we discuss future parity violating measurements of neutron densities.
These purely electro-weak experiments follow in the same tradition and can be
both accurate and model independent.  
Neutron density measurements have implications for nuclear structure, atomic
parity nonconservation (PNC) experiments, isovector interactions, the structure
of neutron rich radioactive beams, and neutron rich matter in astrophysics.  It
is remarkable that a single measurement has so many applications in atomic,
nuclear and astrophysics.

Donnelly, Dubach and Sick\cite{donnelly} suggested that parity violating
electron scattering can measure neutron densities.  This is because the
$Z-$boson couples primarily to the neutron at low $Q^2$.  Therefore one can
deduce the weak-charge density and the closely related neutron density from
measurements of the parity-violating asymmetry in polarized elastic scattering.

Of course the parity violating asymmetry is very small, of order a part per
million.  Therefore measurements were very difficult.  However, a great deal of
experimental progress has been made since the Donnelly {\it et. al.}  
suggestion, and since the early SLAC experiment\cite{SLAC}. 
This includes the Bates $^{12}$C experiment\cite{carbon12}, 
Mainz $^{9}$Be experiment\cite{Heil}, SAMPLE\cite{sample1} 
and HAPPEX\cite{happex1}.  The relative speed of the HAPPEX result and the very 
good helicity correlated beam properties of CEBAF show that very accurate 
parity violation measurements are possible.  Parity violation is now an 
established and powerful tool.

It is important to test the Standard Model at low energies with atomic parity nonconservation (PNC), see for example the Colorado measurement in Cs\cite{Wieman98,Wieman99}.  These
experiments can be sensitive to new parity violating interactions such as
additional heavy $Z-$bosons.  Furthermore, by comparing atomic PNC to higher
$Q^2$ measurements, for example at the $Z$\ pole, one can study the momentum 
dependence of Standard model radiative corrections.  However, as the 
accuracy of atomic PNC experiments improves they
will require increasingly precise information on neutron
densities\cite{Pollock,Chen}.  
This is because the parity violating interaction is
proportional to the overlap between electrons and neutrons.  In the future the
most precise low energy Standard Model test may involve the combination of an
atomic PNC measurement and parity violating electron scattering to constrain the
neutron density.

There have been many measurements of neutron densities with strongly 
interacting probes such as pion or proton elastic scattering, see for 
example ref. \cite{Ray}.  Unfortunately, all such
measurements suffer from potentially serious theoretical systematic errors.  As
a result no hadronic measurement of neutron densities has been generally
accepted by the field.  Because of the uncertain systematic errors, modern mean
field interactions are typically fit without using any neutron density
information.

Finally, there is an important complementarity between neutron radius
measurements in a finite nucleus and measurements of the neutron radius of a
neutron star.  Both provide information on the equation of state of dense
matter.  In a nucleus, $R_n$ is sensitive to the density dependence of the symmetry energy.
Likewise the neutron star radius depends also on the density dependence of the
symmetry energy at normal and somewhat higher densities.
In the future, we expect a number of improving radius measurements for nearby
isolated neutron stars such as Geminga\cite{Geminga} and 
RX J185635-3754\cite{Nature}.

We now present general considerations for neutron density measurements, discusses 
possible theoretical corrections and then conclude.

\section*{General Considerations}

In this section we illustrate how parity violating electron scattering
measures the neutron density and discuss the effects of Coulomb
distortions and other corrections.  These corrections are either small or well known so the interpritation of a measurement is clean.

\subsection*{Born Approximation Assymetry}

The effect of the parity-violating part of the weak interaction may be isolated by measuring the parity-violating asymmetry in the cross section for the scattering of left(right) handed electrons.  In Born approximation the parity-violating asymmetry is,
\begin{equation}A_{LR}=\frac{G_FQ^2}{4\pi\alpha\sqrt{2}}
\Biggl[4\sin^2\theta_W-1+\frac{F_n(Q^2)}{F_p(Q^2)}\Biggr],
\label{equation_bornasy}
\end{equation} 
with $G_F$ the Fermi constant and $\theta_W$ the weak mixing angle.  The Fourier transform of the proton distribution is $F_p(Q^2)$ while that of the neutron distribution is $F_n(Q^2)$ and $Q^2$ is the momentum transfer squared.  The
asymmetry is proportional to $G_F Q^2/\alpha$ which is
just the ratio of $Z^0$ to photon propagators. 
Since 1-4sin$^2\theta_W$ is small and $F_p(Q^2)$ is known we see that $A_{LR}$ directly measures $F_n(Q^2)$.  Therefore, $A_{LR}$ provides a practical method to cleanly measure the
neutron form factor and hence $R_n$.

\subsection*{Coulomb distortions}

By far the largest known correction to the asymmetry comes from coulomb
distortions.  By coulomb distortions we mean repeated electromagnetic
interactions with the nucleus remaining in its ground state.  All of the $Z$
protons in a nucleus can contribute coherently so distortion corrections are
expected to be of order $Z\alpha/\pi$.  This is 20 \% for ${}^{208}$Pb.

Distortion corrections have been accurately calculated in ref.\cite{cjh}.  Here
the Dirac equation was numerically solved for an electron moving in a coulomb
and axial-vector weak potentials.  From the phase shifts, all of the elastic
scattering observables including the asymmetry can be calculated.

Other theoretical corrections from meson exchange currents, parity admixtures in the ground state, dispersion corrections, the neutron electric form factor, strange quarks, the dependence of the extracted radius on the surface shape, etc. are discussed in reference \cite{bigpaper}.  These are all small.  Therefore the interpretation of a parity violating measurement is very clean.

\section*{Conclusion}
With the advent of high quality electron beam facilities
such as CEBAF, experiments for accurately measuring the
weak density in nuclei through parity violating
electron scattering (PVES) are feasible.
The measurements are cleanly interpretable, 
analogous to electromagnetic
scattering for measuring the charge distributions in
elastic scattering.  From parity violating asymmetry measurements
in elastic scattering, one can extract the weak charge density in nuclei and from this the neutron density.

By a direct comparison to theory, these measurements
test mean field theories and other models of the size and shape of nuclei.  
They therefore can have a fundamental and lasting
impact on nuclear physics.  Furthermore, PVES measurements have important implications
for atomic parity nonconservation (PNC) experiments.  In the future it may be possible to combine atomic PNC experiments and PVES to provide a precise test of the Standard Model at low energies.

\section*{Acknowledgments}
This work was done in collaboration with Robert Michaels, Steven Pollock and Paul Souder.  It was supported in part by DOE grant: DE-FG02-87ER40365.

\end{document}